%% file: main.tex
\documentclass[sigplan,screen]{acmart}
\usepackage{balance}
\usepackage[utf8]{inputenc}
\usepackage[T1]{fontenc}
\usepackage{microtype}

\usepackage{booktabs}   
\usepackage{subcaption} 
\usepackage{amssymb}
\usepackage{amsmath}
\usepackage{proof}
\usepackage{relsize}
\usepackage{xcolor}
\usepackage{xspace}
\usepackage{stmaryrd}
\usepackage{tikz}
\usetikzlibrary{shapes,arrows}
\usepackage{pgfplots}
\usepackage{bm}
\usetikzlibrary{shapes.geometric, arrows}
\usetikzlibrary{backgrounds}
\usetikzlibrary{calc}
\usepackage{multirow}
\usepackage[inline]{enumitem}
\usepackage{soul}
\usepackage{mathtools}%
\usepackage{silence}
\usepackage{xtab}

\begin{document}


\title{Multi-layer Optimizations for End-to-End Data Analytics}



 \author{Amir Shaikhha}
 \affiliation{
   \institution{University of Oxford}
   \country{United Kingdom}
 }

 \author{Maximilian Schleich}
 \affiliation{
   \institution{University of Oxford}
  \country{United Kingdom}
 }

 \author{Alexandru Ghita}
 \affiliation{
   \institution{University of Oxford}
  \country{United Kingdom}
 }

 \author{Dan Olteanu}
 \affiliation{
   \institution{University of Oxford}
  \country{United Kingdom}
 }

\renewcommand{\shortauthors}{}

\setcopyright{none}
\renewcommand\footnotetextcopyrightpermission[1]{}
\settopmatter{printacmref=false}
\pagestyle{plain} 

\begin{abstract}
We consider the problem of training machine learning models over multi-relational data. The mainstream approach is to first construct the training dataset using a feature extraction query over input database and then use a statistical software package of choice to train the model. In this paper we introduce Iterative Functional Aggregate Queries (IFAQ), a framework that realizes an alternative approach. IFAQ treats the feature extraction query and the learning task as one program given in the IFAQ's domain-specific language, which captures a subset of Python commonly used in Jupyter notebooks for rapid prototyping of machine learning applications. The program is subject to several layers of IFAQ optimizations, such as algebraic transformations, loop transformations, schema specialization, data layout optimizations, and finally compilation into efficient low-level C++ code specialized for the given workload and data.

We show that a Scala implementation of IFAQ can outperform mlpack, Scikit, and TensorFlow by several orders of magnitude for linear regression and regression tree models over several relational datasets.
\end{abstract}





\maketitle
\input{macros}

\input{intro}

\input{arch}

\input{apps}

\input{opts}

\input{exp}

\input{related}

\balance

\input{concl}

\paragraph{Acknowledgments} This project has received funding from the European Union's Horizon 2020 research and innovation programme under grant agreement No 682588.

\bibliographystyle{ACM-Reference-Format}
\bibliography{main}
\end{document}

%% file: macros.tex
\newcommand{\smartpara}[1]{\noindent \textbf{#1.}}

\newcommand{\lang}{IFAQ\xspace}
\newcommand{\dlang}{D-IFAQ\xspace}
\newcommand{\slang}{S-IFAQ\xspace}
\newcommand{\system}{IFAQ\xspace}

\newcommand{\code}[1]{\texttt{#1}}
\newcommand{\variant}[1]{\code{<}#1\code{>}}
\newcommand{\record}[1]{\code{\{}#1\code{\}}}
\newcommand{\bag}[1]{{\code{Bag[}#1\code{]}}}
\newcommand{\maptype}[2]{{\code{Map[}#1\code{, }#2\code{]}}}
\newcommand{\gbag}[2]{\maptype{#1}{#2}}
\newcommand{\indextype}{\text{$\mathbb{N}$}}
\newcommand{\realtype}{\text{$\mathbb{R}$}}
\newcommand{\settype}[1]{{\code{Set[}#1\code{]}}}
\newcommand{\hottype}[2]{\text{$\mathbb{R}^{\text{#1}}_{\text{#2}}$}}
\newcommand{\while}{\code{while}}
\newcommand{\sspace}{\text{ }}
\newcommand{\return}{\sspace}
\newcommand{\convloop}[3]{\code{iterate(}#1\code{, }#2\code{, }#3\code{)}}
\newcommand{\assign}{\text{$\leftarrow$}}
\newcommand{\fullconvloop}[4]{\text{#1 \assign #2 \while\texttt{(}#3\texttt{) \{} #1 \assign #4\texttt{\}} \return #1}}
\newcommand{\lett}{\code{let }}
\newcommand{\inn}{\code{in }}
\newcommand{\letbinding}[3]{\lett #1 = #2\sspace\inn #3}
\newcommand{\letbind}[2]{\lett #1 = #2 \code{in}}
\newcommand{\ifthenelse}[3]{\code{if } #1  \code{ then } #2 \code{ else } #3}
\newcommand{\fieldreflect}[2]{#1\code{[}#2\code{]}}
\newcommand{\grammarcomment}[1]{}
\newcommand{\set}[1]{\code{[\![}#1\code{]\!]}}
\newcommand{\dictmapping}{\text{$\rightarrow$}}
\newcommand{\dictmult}[1]{\code{\{\!\{}#1\code{\}\!\}}}
\newcommand{\dict}[2]{\dictmult{#1 \dictmapping #2}}
\newcommand{\dom}[1]{\code{dom(}#1\code{)}}
\newcommand{\fieldtype}{\code{Field}}
\newcommand{\fieldlit}[1]{\code{`}#1\code{`}}
\newcommand{\stringlit}[1]{\code{"}#1\code{"}}

\newcommand{\codespace}{}

\newcommand{\todo}[1]{\textcolor{red}{#1}}

\newcommand{\evalsto}{\text{ $\leadsto$ }}
\newcommand{\contexthole}[1]{\textbf{$\bm\Gamma$(}#1\textbf{)}}

\newcommand{\expr}{\text{e}}
\newcommand{\gind}[2]{\text{$\text{#1}_{#2}$}}
\newcommand{\exprind}[1]{\gind{\expr}{#1}}
\newcommand{\mult}{\text{$*$}\xspace}
\newcommand{\add}{\text{$+$}\xspace}
\newcommand{\rows}[1]{\code{rs(}#1\code{)}}
\newcommand{\cols}[1]{\code{cs(}#1\code{)}}
\def\tabt{\hspace*{0.3cm}}
\newcommand{\samerule}{}
\newcommand{\nextrule}{}

\tikzstyle{startstop} = [rectangle, rounded corners, minimum width=1.5cm, minimum height=1cm,text centered, draw=black, fill=white!30, scale=0.9]

\tikzstyle{arrow} = [thick,->,>=stealth]

\newcommand{\gv}[1]{\ensuremath{\mbox{\boldmath$ #1 $}}} 

\newcommand{\whilecond}{\code{ not converged }}

\newcommand{\norm}[1]{\left\|#1\right\|}
\newcommand{\grad}[1]{\gv{\nabla} #1} 
\newcommand{\mv}[1]{\mathbf{#1}}
\newcommand{\inner}[1]{\left\langle #1 \right\rangle}
\newcommand{\pd}[2]{\frac{\partial#1}{\partial#2}}
\renewcommand{\vec}[1]{\ensuremath\boldsymbol{#1}}
\newcommand{\col}[1]{\textbf{#1}}
\newcommand{\colm}[1]{\bm #1}
\newcommand{\colind}[2]{\col{\gind{#1}{#2}}}

\colorlet{dred}{red!80!black}
\colorlet{dgreen}{green!50!black}
\definecolor{dblue}{rgb}{0, 0.33, 0.71}

\newcommand{\multicolor}[2]{{\leavevmode\color{#1}[#2]}}

\newcommand{\spcfigb}{\vspace{-0.5cm}}

\makeatletter
\DeclareRobustCommand\bigop[1]{%
  \mathop{\vphantom{\sum}\mathpalette\bigop@{#1}}\slimits@
}
\newcommand{\bigop@}[2]{%
  \vcenter{%
    \sbox\z@{$#1\sum$}%
    \hbox{\resizebox{\ifx#1\displaystyle.9\fi\dimexpr\ht\z@+\dp\z@}{!}{$\m@th#2$}}%
  }%
}
\makeatother

\newcommand{\biglam}{\DOTSB\bigop{\lambda}}
\newcommand{\bigsum}{\DOTSB\bigop{\Sigma}}

\newcommand{\dictbuild}[2]{\text{$\biglam\limits_{#1}#2$}}
\newcommand{\summation}[2]{\text{$\bigsum\limits_{#1}#2$}}
\definecolor{gray1}{gray}{0.1}
\definecolor{gray2}{gray}{0.2}
\definecolor{gray3}{gray}{0.3}
\definecolor{gray4}{gray}{0.4}
\definecolor{gray5}{gray}{0.5}

%% file: intro.tex
\section{Introduction}
The mainstream approach in supervised machine learning over relational data 
is to specify the construction of the data matrix followed by the learning task in
scripting languages such as Python, MATLAB, Julia, or R using software
environments such as Jupyter notebook. These environments call libraries for
query processing, e.g., Pandas~\cite{mckinney2011pandas}
or SparkSQL~\cite{Armbrust:2015:SSR:2723372.2742797}, or  database
systems, e.g., PostegreSQL~\cite{momjian2001postgresql}.
The materialized training dataset then becomes the input to a statistical package, e.g., mlpack~\cite{mlpack2018}, scikit-learn~\cite{pedregosa2011scikit}, TensorFlow~\cite{abadi2016tensorflow}, and PyTorch~\cite{paszke2017automatic}) that learns the model. 
There are clear advantages to this approach: 
it is easy to use, allows for quick prototyping, and does
not require substantial programming knowledge.
Although it uses libraries that provide efficient implementations 
for their functions (usually by binding to efficient low-level C code), 
they miss optimization opportunities for the end-to-end relational learning pipeline. 
In particular, they fail to exploit the relational structure of the underlying data, 
which was removed by the materialization of the training dataset. 
The effect is that the efficiency on one machine is often severely limited, with common deployments having to rely on expensive and energy-consuming clusters of machines to perform the learning task. 

The thesis of this paper is that this performance limitation can be overcome by systematically optimizing the end-to-end relational learning pipeline as a whole. 
We introduce Iterative Functional Aggregate Queries, or \lang for short, a framework that is designed to uniformly process and automatically optimize the various phases of the relational learning pipeline. \lang takes as input a program that fully specifies both the data matrix construction and the model training in a dynamically-typed language called \dlang. This language captures a fragment of scripting languages such as Python that is used for rapid prototyping of machine learning models. This is possible thanks to the iterative and collection-oriented constructs provided by \dlang.

An \lang program is optimized by multiple compilation layers, whose optimizations are inspired by techniques developed by the data management, programming languages, and high-performance computing communities. 
\lang performs automatic memoization, which identifies code fragments that can be expressed as batches of aggregate queries. This optimization further enables non-trivial loop-invariant code motion opportunities. 
The program is then compiled down to a statically-typed language, called \slang,
which benefits from further optimization stages, including loop transformations such as loop fusion and code motion. 
\lang investigates optimization opportunities at the interface between the data matrix construction and learning steps and interleaves the code for both steps by pushing the data-intensive computation from the second step past the joins of the first step, inspired by recent query processing techniques~\cite{Olteanu:2016:FD:3003665.3003667,abo2016faq}.
The outcome is a highly optimized code with no separation between query processing and machine learning.
Finally, \lang compiles the optimized program into low-level C++ code that further exploits data-layout optimizations. \lang can outperform equivalent solutions by orders of magnitude.

The contributions of this paper are as follows:
\begin{itemize}[leftmargin=.1in]
\item Section~\ref{sec:overview} introduces the \lang framework, which comprises languages and compiler optimizations that are designed for efficient end-to-end relational learning pipelines.
\item As proof of concept, Section~\ref{sec:apps} demonstrates \lang for two popular models: linear regression and regression tree.
\item Section~\ref{sec:opts} shows how to systematically optimize an \lang program in several stages.
\item Section~\ref{sec:exp} benchmarks \lang, mlpack, TensorFlow, and scikit-learn. 
It shows that \lang can train linear regression and regression tree models over two real datasets orders-of-magnitude faster than its competitors. In particular, \lang learns the models faster than it takes the competitors to materialize the training dataset.
  Section~\ref{sec:exp} further shows the performance impact of individual optimization layers. 
\end{itemize}


%% file: arch.tex
\input{figures/arch-fig}

\section{Overview}
\label{sec:overview}

Figure~\ref{fig:overview} depicts the \lang workflow. Data scientists use software environments such as Jupyter Notebook to specify relational learning programs. In our setting, such programs are written in a dynamically-typed language called \dlang and subject to high-level optimizations (Section~\ref{sec:hlopt}). Given the schema information of the dataset, the optimized program is translated  into a statically-typed language called \slang (Section~\ref{sec:schemaspec}). If there are type errors, they are reported to the user. IFAQ performs several optimizations inspired by database query processing techniques (Section~\ref{sec:lmfaoopt}).  Finally, it synthesizes appropriate data layouts, resulting in efficient low-level C/C++ code (Section~\ref{sec:dsopt}).

\subsection{IFAQ Core Language}
\label{sec:lang}
The grammar of the \lang core language is given in Figure~\ref{fig:lang}.
This functional language support the following data types. The first category consists of numeric types as well as
categorical types (e.g., boolean values, string values, and other custom enum types). 
Furthermore, for categorical types the other alternative is to one-hot encode them,  
the type of which is represented as \hottype{$n$}{$T_1$}.
This type represents an array of $n$ real numbers
each one corresponding to a particular value in the domain of $T_1$.
In the one-hot encoding of a value of type $T_1$, only the $i^{th}$ value is 1 and the rest
are zero. However, a value of type 
 denoting the one-hot encoding of \hottype{$n$}{$T_1$}, can take arbitrary real numbers at each position.

The second category consists of record types, which are similar to \textit{struct}s in C;
the record values contain various fields of possibly different data types. 
The partial records, where some fields have no values are referred to 
as \textit{variants}.

\input{figures/lang}

The final category consists of collection data types such as (ordered) sets and dictionaries. Database relations are represented as dictionaries 
(mapping elements to their multiplicities).
In \dlang, the elements of collections can have different types.
However, in \slang the elements of collections should have the same data type.
In order to distinguish the variables with collection data type with other types of
variables, we denote them as \col{x} and $x$, respectively.

The top-level program
consists of several initialization expressions, followed by a loop. This enables \lang 
to express iterative algorithms such as gradient descent optimization algorithms.  
The rest of the constructs of this language are as follows:
\begin{enumerate*}[label=(\roman*)]
\item various binary and unary operations on both scalar and collection data structures,\footnote{More specifically ring-based operations.}
\item let binding for assigning the value of an expression to a variable,
\item conditional expressions
\item $\sum\nolimits_{x \in \exprind{1}}\exprind{2}$: the summation operator in order to iterate over the elements of a 
collection (\exprind{1}) and perform a stateful computation using a particular addition operator 
(e.g., numerical addition, set union, bag union),\footnote{This operator could be every monoid-based operator. 
Thus, even computing the minimum of two numbers is an addition operator.}
\item
  $\operatorname*{{\mathlarger{\lambda}}}\nolimits_{x \in
    \exprind{1}}\exprind{2}$: constructing a dictionary where the key domain is
  \exprind{1} and the value for each key $x$ is constructed using \exprind{2},
\item constructing sets (\set{\expr}) and dictionaries (\dict{\expr}{\expr}) given a list of elements,
\item \dom{\expr}: getting the domain of a dictionary,
\item \exprind{0}({\exprind{1}}): retrieving the associatiated value to the given key in a dictionary,
\item constructing records ($\record{\overrightarrow{x=e}}$) and variants ($\variant{x=e}$),
\item statically (\expr.f) and dynamically (\fieldreflect{\expr}{f}) assessing the field of a record or a variant.
\end{enumerate*}



%% file: figures/arch-fig.tex
\begin{figure}[t]
\begin{tikzpicture}[node distance=1.8cm]

\node (notebook) [text width=1.3cm, align=center] {\footnotesize Relational\\ \footnotesize Learning \\\footnotesize Task};
\node (difaq) [startstop, right of=notebook, align=center, xshift=0.1cm] {\dlang};
\node (sifaq) [startstop, right of=difaq, align=center] {\slang};
\node (ccode) [startstop, right of=sifaq, align=center] {C/C++};

\draw [arrow] (notebook) -- node[align=center, anchor=east] {} (difaq);
\draw [arrow] (difaq) -- node[align=center, anchor=east] {} (sifaq);
\draw [arrow] (sifaq) -- node[align=center, anchor=east] {} (ccode);

\draw [dashed] (0.93,0.5) rectangle (4.1,-0.5);

\node[draw=none] at (2.5,-0.9) {\textbf{IFAQ}};

\end{tikzpicture}
\vspace{-.25cm}
\Description{A relational learning task is expressed in \dlang, transformed into an optimized \slang expression, and compiled to efficient C++ code.}
\caption{A relational learning task is expressed in \dlang, transformed into an optimized \slang expression, and compiled to efficient C++ code.}
\label{fig:overview}
\spcfigb
\end{figure}
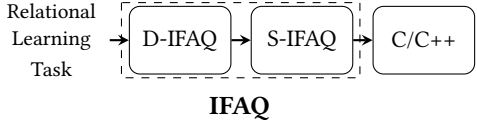

%% file: figures/lang.tex
\begin{figure}[t]
\begin{eqnarray*}
p    &\mbox{::=}& 
e \mid \fullconvloop{x}{e}{e}{e}\\
e    &\mbox{::=}& e+e \mid e*e \mid -e \grammarcomment{Ring operations} 
     \mid uop(e) \mid e ~ bop ~ e \mid c\\
     &\mid& \summation{x \in e}{e} \grammarcomment{Summation over a set} 
     \mid \dictbuild{x \in e}{e} 
     \mid \dictmult{\overrightarrow{e \rightarrow e}} \grammarcomment{Dictionary constructor} 
     \mid \set{\overrightarrow{e}} \grammarcomment{Set constructor} 
     \mid \dom{e} \grammarcomment{Key Set of a Dictionary}
     \mid e(e) \grammarcomment{Accessing an Element of a Dictionary}\\
     &\mid& \record{\overrightarrow{x=e}} \mid \variant{x=e}
     \mid e.x \mid \fieldreflect{e}{e} \grammarcomment{Static \& Reflective Field Access}\\
     &\mid& \letbinding{x}{e}{e} \mid x 
     \mid \ifthenelse{e}{e}{e} \\
c    &\mbox{::=}& \fieldlit{id} \mid \stringlit{id} \mid n \mid r \mid \code{false} \mid \code{true} \\
T    &\mbox{::=}& S \mid \record{\overrightarrow{x:T}} \mid \variant{\overrightarrow{x:T}} \mid \gbag{T}{T} \mid \settype{T} \\
S    &\mbox{::=}& B \mid C \\
C    &\mbox{::=}& \code{string} \mid \code{enum} \mid \code{bool} \mid \fieldtype \\
B    &\mbox{::=}& \mathbb{Z} \mid \mathbb{R} \mid \hottype{n}{C}
\end{eqnarray*}
\vspace{-.7cm}
\Description{Grammar of \lang core language.}
\caption{Grammar of \lang core language.}
\label{fig:lang}
\spcfigb
\end{figure}

%% file: apps.tex
\section{Applications}
\label{sec:apps}

As proof of concept, we show in this section how linear regression and regression tree models can be expressed in \dlang. We assume that the model is learned over a training dataset \col{Q} with features $a_1,\ldots, a_n$ and label $a_{n+1}$, which is the output of a query over a database with relations $\colm{R_1}, \ldots, \colm{R_m}$.

Further applications that can benefit from \lang include supervised learning problems (e.g., logistic regression, factorization machines, and support vector machines), as well as unsupervised learning problems (e.g., k-means clustering, principal component analysis, and learning Bayesian networks via Chow-Liu trees). \lang further supports different optimization algorithms, e.g., batch or stochastic gradient descent, coordinate descent, and alternating minimization.

\noindent{\bf Linear Regression:} A linear regression model is given by:
\begin{small}
  \begin{align*}
    LR(x) &= \sum\nolimits_{f \in \col{F}} \colm{\theta}(f) \mult x[f]
  \end{align*}
\end{small}
where $x$ is a record with fields $\col{F} = \{a_1,\ldots, a_n\}$, and
$\colm{\theta}$ is a dictionary with these fields as its keys. The
$\colm{\theta}$ dictionary defines the parameters of the model, and the output
of $LR(x)$ is a prediction for the label given input $x$.

Linear regression models require categorical features to be one-hot
encoded. Thus, for a categorical feature $a_i$, the field $x[a_i]$ encodes a
vector of one indicator value for each value in the domain of $a_i$
(Section~\ref{sec:lang}).  The corresponding parameter element
$\colm{\theta}(a_i)$ encode a vector of parameters which is of the same
dimension as $x[a_i]$. We assume without loss of generality that $x[a_1]$ only
takes value $1$ and then $\colm{\theta}(a_1)$ is the intercept of the model.

Given the training dataset \col{Q}, the error of fitting $LR$ to \col{Q} is given by the sum of the least squares loss function:\footnote{It is common to penalize
  $J(\colm{\theta})$ with a regularization term. \lang supports regularization
  as well, but we omit it from the examples to avoid clutter.}
\begin{small}
  \begin{align*}
    J(\colm{\theta})
    &= \frac{1}{2|\col{Q}|}  \sum_{x \in \dom{\col{Q}}} \big(
      \sum_{f \in \col{F}} \colm{\theta}(f)\mult x[f] - x[a_{n+1}]\big)^2 
  \end{align*}
\end{small}
We minimize $J(\colm{\theta})$ using batch gradient descent (BGD) optimization,
which repeatedly updates each parameter $\colm{\theta}(f)$ by learning rate
$\alpha$ in the direction of the partial derivative of $J(\colm{\theta})$
w.r.t. $\colm{\theta}(f)$ until convergence. This is represented in the
following \dlang program (\col{Q} is expressed in \system as a query over a multi-relational database):
\begin{small}
\begin{align*}
  &\letbinding{\col{F}}{\set{a_1,a_2,\ldots,a_n}}{}\\
  &\colm{\theta} \leftarrow \colm{\theta_0}\\
  &\code{while(}\whilecond{}\code{)\;\{}\\
  &\quad\colm{\theta} = \dictbuild{f_1 \in \col{F}}{}\Big(\colm{\theta}(f_1) - {} \\
    &\hspace{3.7em}\frac{\alpha}{|\col{Q}|}\summation{x \in \dom{\col{Q}}}{} \hspace{-0.5em}\col{Q}(x) *(\summation{f_2 \in \col{F}}{}
    \colm{\theta}(f_2)  * x[f_2] - x[a_{n+1}])* x[f_1]\Big)\\[-0.2em]
  &\code{\}}\\
  &\colm{\theta}
\end{align*}
\end{small}
\noindent This program is inefficient, because it computes for each BGD
iteration all pairwise products $x[f_1] * x[f_2]$. These products, however, are
invariant, and can be computed once outside the while loop. This is the main
high-level optimization performed by \lang for this case (Section~\ref{sec:hlopt}). The rewriting would
gather a collection of these products, which can be computed once as database
aggregate queries. The following BGD iterations can then be computed directly
over these aggregates, and do not require a pass over $\col{Q}$. The collection
of these aggregates represents the \textit{non-centered covariance matrix} (or
\textit{covar matrix} for short).  The aggregates can then be further optimized
using ideas from the database query processing, in particular they can be
computed directly over the input database and without materializing
$\col{Q}$ (Section~\ref{sec:lmfaoopt}).

We next provide a running example which is used to explain the optimization
rewritings in the following sections.

\begin{example}
	In retail forecasting scenarios, the goal is to
learn a model which can predict future sales for items at different stores. We
assume the retailer has a database with three relations:
\col{S}ales(\textit{\underline{i}tem, \underline{s}tore, \underline{u}nits}),
Sto\col{R}es(\textit{\underline{s}tore, \underline{c}ity}),
\col{I}tems(\textit{\underline{i}tem, \underline{p}rice}).  The goal is to learn
a model that predicts $u$ with the features $\col{F} = \{i, s, c, p\}$.  The training dataset is given by the
result of the join of the three relations:
$\col{Q}=\col{S}\bowtie \col{R}\bowtie \col{I}$.

We learn the model with BGD using the \dlang program above. To avoid clutter, we
focus on the core computation of the program and we make two simplifications:
(1) we assume $\frac{\alpha}{|\col{Q}|} = 1$, and (2) we hide the term for
$x[a_{n+1}]$. Then the inner-loop expression is given by:

\noindent
\begin{small}
\begin{align*}
  &\colm{\theta} = \dictbuild{f_1 \in \col{F}}{}\Big( \colm{\theta}(f_1)\, -
    \hspace{-0.5em}\summation{x \in \dom{\col{Q}}}{} \col{Q}(x) * \big(\summation{f_2 \in \col{F}}{}
    \colm{\theta}(f_2) *x[f_2]\big) * x[f_1]\Big)
\end{align*}
\end{small}

\noindent where $\col{F} = \set{\fieldlit{i}, \fieldlit{s}, \fieldlit{c}, \fieldlit{p}}$.
\end{example}

\input{figures/lang-opts-fig}

\noindent{\bf Decision Tree.}
We next consider learning decision trees using the CART
algorithm~\cite{breiman1984classification}.  With our optimizations, we can
learn both classification and regression trees efficiently. In the following, we
focus on regression trees.

A regression tree model is a (binary) tree with inner nodes representing
conditional control statements to model decisions and their consequences.
Given an element $x$ in the dataset \col{Q}, the condition for a feature $f$ is of 
the form $x[f] \text{ op } t$, which is denoted as $c(f, \text{op}, t)$ here.  For
categorical features (e.g., city), $t$ may be a set of categories and
$\text{op}$ denotes inclusion. For continuous features (e.g., price), $t$ is a
real number and $\text{op}$ is inequality. Leaf nodes represent predictions for
the label. For regression trees, the prediction is the average of the label
values in the fragment of the training dataset that satisfies all control
statements on the root to leaf path.

For a given node $N$, let $\delta$ encode
the conjunction of all conditions along the path from the root to $N$. The CART algorithm recursively seeks for the condition 
that minimises the following optimization problem for a given cost function: 
\begin{align*}
\min_{f \in \colm{F}}  \min_{t \in \colm{T_f}} \texttt{cost}(\col{Q}, \delta \land c(f, \text{op}, t)) + \texttt{cost}(\col{Q}, \delta \land c(f, \text{!op}, t)). 
\end{align*}
where \texttt{!op} denotes the negation of \texttt{op}, $\colm{T_f}$ is the set
of all possible thresholds for $f$.
Once this condition is found, a new node with condition $c(f, \text{op}, t)$ is
  constructed and the algorithm recursively constructs the next node
  for each child.

For regression trees, the cost is given by the variance, which is represented as the following \dlang expression:

\noindent
\begin{small}
\begin{align*}
    \texttt{cost}(\col{Q}, \delta')=&
    \sum\limits_{x\in  \dom{\col{Q}}}\col{Q}(x)\mult{}x[a_{n+1}]^2\mult{}\delta'-\\&\frac{1}{\sum\limits_{x\in  \dom{\col{Q}}}\col{Q}(x)\mult{}\delta'}(
     \sum\limits_{x\in \dom{\col{Q}}}\col{Q}(x)\mult x[a_{n+1}]\mult{}\delta')^2
\end{align*}
\end{small}

\noindent In this formula, $\delta'$ can be $\delta \land c(f, \text{op}, t)$ or $\delta \land c(f, \text{!op}, t)$.


In contrast to linear regression, all aggregates depend on
node-specific information ($\delta'$). It is thus not possible to
hoist and compute them only once for all recursions of the CART
algorithm. Nevertheless, all other optimizations presented in the next section are applicable.


%% file: figures/lang-opts-fig.tex
\begin{figure*}[t!]
\centering
\tikzstyle{block} =[rectangle, minimum width=4.5cm, minimum height=0.55cm, text centered, text width=4cm, draw=black, rounded corners]
\tikzstyle{trans} =[rectangle, minimum width=2.2cm, minimum height=0.2cm, text centered, text width=4.8cm, draw=white, rounded corners, fill=black, text=white, yshift=0.05cm, scale=0.6]
\tikzstyle{nblock} =[rectangle, minimum width=2.2cm, minimum height=0.55cm, text centered, text width=1.8cm, draw=black, rounded corners]
\tikzstyle{ablock} =[rectangle, minimum width=1.7cm, minimum height=0.95cm, text centered, text width=1.4cm, draw=black, rounded corners]
\tikzstyle{sblock} =[rectangle, minimum width=1.7cm, minimum height=0.55cm, text centered, text width=1.4cm, draw=black, rounded corners, fill=lightgray]
\tikzstyle{oblock} =[rectangle, minimum width=1.7cm, minimum height=1.1cm, text centered, text width=1.4cm, draw=black]
\tikzstyle{anblock} =[circle, minimum width=1.5cm, minimum height=1.1cm, text centered, text width=1.2cm, draw=black]
\tikzstyle{fp} =[rectangle, minimum width=0.5cm, minimum height=0.55cm, fill=blue]
\tikzstyle{ip} =[rectangle, minimum width=0.5cm, minimum height=0.55cm, fill=red]
\tikzstyle{line} = [draw, -latex']
\tikzstyle{plus} =[rectangle, text width=2.4cm]
\newcommand{\schappdist}{1.75cm}
\newcommand{\sblockdist}{2.45cm}
\begin{tikzpicture}[node distance = 2.5cm, auto]
   \node[ablock](app1){\dlang Expr.};
   \node[sblock, below of=app1,yshift=\schappdist](sch1){\footnotesize Schema};
   \node[oblock, right of=app1](opt1){\tiny High-Level Optimizations};
   \node[ablock, right of=opt1](app2){\dlang Expr.};
   \node[sblock, below of=app2,yshift=\schappdist](sch2){\footnotesize Schema};
   \node[oblock, right of=app2](opt2){\tiny Schema Specialization};
   \node[ablock, right of=opt2](app3){\slang Expr.};
   \node[sblock, below of=app3,yshift=\schappdist](sch3){\footnotesize Schema};
   \node[anblock, right of=app3](qopt1){\tiny Extract Aggregates};
   \node[sblock, right of=qopt1,yshift=0.1cm](aggs){};
   \node[sblock, below of=aggs,xshift=-0.05cm,yshift=\sblockdist](aggsc1){};
   \node[sblock, below of=aggsc1,xshift=-0.05cm,yshift=\sblockdist](aggsc2){\footnotesize Aggregates};
   \node[anblock, below of=aggs,yshift=0.5cm](qopt2){\tiny Join Tree Construction};
   \node[sblock, left of=qopt2](vt4){\footnotesize Join Tree};
   \node[oblock, left of=vt4](qopt4){\tiny Aggregate Push Down};
   \node[ablock, left of=qopt4,yshift=0.75cm](app5){\slang Expr.};
   \node[sblock, below of=app5,yshift=\schappdist](sch5){\footnotesize Schema};
   \node[sblock, below of=sch5,yshift=1.95cm](vt5){\footnotesize View Tree};
   \node[oblock, left of=sch5](qopt5){\tiny Merge Views};
   \node[ablock, left of=qopt5,yshift=0.75cm](app6){\slang Expr.};
   \node[sblock, below of=app6,yshift=\schappdist](sch6){\footnotesize Schema};
   \node[sblock, below of=sch6,yshift=1.95cm](vt6){\footnotesize View Tree};
   \node[oblock, below of=vt6,yshift=1.15cm](qopt7){\tiny Multi-Aggregate Iteration};
   \node[ablock, right of=qopt7,yshift=0.7cm](app7){\slang Expr.};
   \node[sblock, below of=app7,yshift=\schappdist](sch7){\footnotesize Schema};
   \node[oblock, right of=sch7](dsopt){\tiny Data-Layout Synthesis};
   \node[ablock, right of=dsopt](app8){Expr. in C/C++};

   \node[ablock, right of=app8,xshift=-1.1cm,yshift=0.6cm, minimum width=0.4cm,minimum height=0.4cm,text width=0.2cm](leg1){};
   \node[right of=leg1,xshift=-1.6cm, align=left](leg1text){\small Program};
   \node[sblock, below of=leg1,yshift=1.80cm, minimum width=0.4cm,minimum height=0.4cm,text width=0.2cm](leg2){};
   \node[right of=leg2,xshift=-1.48cm, align=left]{\small Meta-Data};
   \node[oblock, right of=leg1text, minimum width=0.4cm,minimum height=0.4cm,text width=0.2cm,xshift=-1.3cm](leg3){};
   \node[right of=leg3,xshift=-1.15cm, align=left]{\small Program \\ \small Transformation};
   \node[anblock, below of=leg3,yshift=1.80cm, minimum width=0.4cm,minimum height=0.4cm,text width=0.2cm](leg4){};
   \node[right of=leg4,xshift=-1cm, align=left]{\small Program Analysis};
   \begin{scope}[on background layer]
   \draw [arrow] (app1) -- (opt1);
   \draw [arrow] (opt1) -- (app2);
   \draw [arrow] (app2) -- (opt2);
   \draw [arrow] (opt2) -- (app3);
   \draw [arrow] (app3) -- (qopt1);
   \draw [arrow] (qopt1) -- (aggsc2);
   \draw [arrow] (aggs) -- (qopt2);
   \draw [arrow] (qopt2) -- (vt4);
   \draw [arrow] (sch3) -- (qopt4);
   \draw [arrow] (vt4) -- (qopt4);
   \draw [arrow] (qopt4) -- (sch5);
   \draw [arrow] (sch5) -- (qopt5);
   \draw [arrow] (qopt5) -- (sch6);
   \draw [arrow] (vt6) -- (qopt7);
   \draw [arrow] (qopt7) -- (sch7);
   \draw [arrow] (sch7) -- (dsopt);
   \draw [arrow] (dsopt) -- (app8);
   \draw [dashed] (11,-2.7) rectangle (16.3,-4.35);
   \end{scope}
\end{tikzpicture}
\Description{Transformation steps for an expression written in \lang.}
\caption{Transformation steps for an expression written in \lang.}
\label{fig:allopts}

\end{figure*}
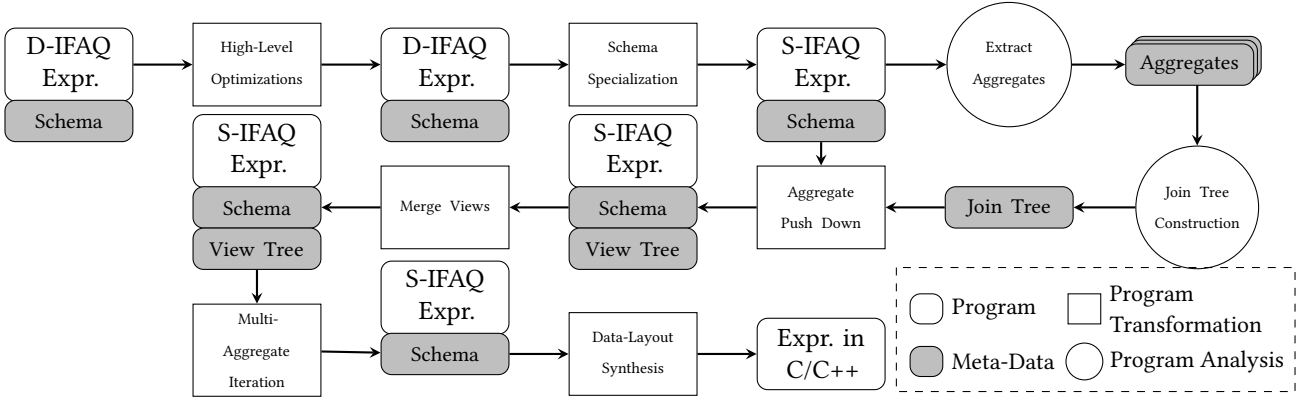

%% file: opts.tex
\section{Optimizations}
\label{sec:opts}

This section details the \system optimizations.
Figure~\ref{fig:allopts} overviews the transformations applied at 
different stages.

\subsection{High-Level Optimizations}
\label{sec:hlopt}

The high-level optimizations applied to the \dlang expressions are: normalization, loop scheduling, factorization, static memoization, and loop-invariant code motion. The first two transformations are preprocessing steps for factorization. The impact of static memoization becomes positive once it is combined with loop-invariant code motion.

\smartpara{Normalization} This transformation brings the expressions into a normalized form. 
Similar to other algebraic-based systems (e.g., logical circuits) the normal form is 
sum of products. Thus, this transformation applies distributivity and pushes the 
products inside summation (Figure~\ref{fig:lang_opt_norm}).

\begin{example}
	The inner-loop of our running example:
\codespace{}

\noindent
\begin{tabular}{l}
$\colm{\theta} = \dictbuild{f_1 \in \col{F}}{}\big(\colm{\theta}(f_1) - \summation{x \in \dom{\col{Q}}}{}(\col{Q}(x) * \summation{f_2 \in \col{F}}{}\colm{\theta}(f_2) * x[f_2]) * x[f_1]\big)$
\end{tabular}

is normalized by pushing $x[f_1]$ into the sum over $f_2$: 

\noindent
\begin{tabular}{l}
$\colm{\theta} = \dictbuild{f_1 \in \col{F}}{}\big(\colm{\theta}(f_1) - \summation{x \in \dom{\col{Q}}}{}\col{Q}(x) * \summation{f_2 \in \col{F}}{}\colm{\theta}(f_2) * x[f_2] * x[f_1]\big)$
\end{tabular}
\end{example}

\smartpara{Loop Scheduling} This transformation reorders nested loops (i.e., nested
summations) such that the outer loop 
iterates over a smaller collection (Figure~\ref{fig:lang_opt_loopsch}).
Pushing larger loops inside allows to factorize the
free variables of the outer smaller loop outside the more expensive inner loop.

\begin{example}
 We next reorder the loops:
\codespace{}

\noindent
\begin{tabular}{l}
$\colm{\theta} = \dictbuild{f_1 \in \col{F}}{}\big(\colm{\theta}(f_1) - \summation{f_2 \in \col{F}}{}\;\summation{x \in \dom{\col{Q}}}{}\col{Q}(x) * \colm{\theta}(f_2) * x[f_2] * x[f_1]\big)$
\end{tabular}
\end{example}

\smartpara{Factorization} Now that the loops are correctly ordered, 
we factorize the multiplication operations outside summation (Figure~\ref{fig:lang_opt_fact}) so that
they are no longer performed in an expensive loop.
This results in less arithmetic operations, which in turn leads to better performance.

\begin{example}
	We next factorize the inner-loop expression:
\codespace{}

\noindent
\begin{tabular}{l}
$\colm{\theta} = \dictbuild{f_1 \in \col{F}}{}\big(\colm{\theta}(f_1) - \summation{f_2 \in \col{F}}{}\colm{\theta}(f_2) * \summation{x \in \dom{\col{Q}}}{}\col{Q}(x) * x[f_2] * x[f_1]\big)$
\end{tabular}
\end{example}

\smartpara{Static Memoization} There are repetitive computations inside a loop that cannot be easily 
hoisted outside. This is because they depend on variables defined inside the loop.
One case stands out: the free variables are bound in loops ranging over a statically-known finite domain. 
In this case we can \textit{memoize} all values as a dictionary mapping from the domain of the free 
variables to the values (Figure~\ref{fig:lang_opt_statmemo}).

\begin{example}
	Static memoization results in:
\codespace{}

\noindent
\begin{tabular}{l}
$\letbinding{\col{M}}{\dictbuild{f_1 \in \col{F}}{}\ \dictbuild{f_2 \in \col{F}}{}\ \summation{x \in \dom{\col{Q}}}{} \col{Q}(x) * x[f_2] * x[f_1] }{}$\\
$\colm{\theta} = \dictbuild{f_1 \in \col{F}}{}\big(\colm{\theta}(f_1) - \summation{f_2 \in \col{F}}{}\colm{\theta}(f_2) * \col{M}(f_2)(f_1)\big)$
\end{tabular}
\end{example}

\smartpara{Loop-Invariant Code Motion} After static memoization, the computations that can be 
shared are no longer dependent on the variables defined in the loop and can 
 be hoisted outside of the loop: A loop-invariant let binding
is moved outside the loop (Figure~\ref{fig:lang_opt_licm}).

\begin{example}
The result of loop-invariant code motion is:
\codespace{}

\noindent
\begin{tabular}{l}
$\letbinding{\col{F}}{\set{\fieldlit{i},\fieldlit{s},\fieldlit{c},\fieldlit{p}}}{}$\\
$\letbinding{\col{M}}{\dictbuild{f_1 \in \col{F}}{}\ \dictbuild{f_2 \in \col{F}}{}\ \summation{x \in \dom{\col{Q}}}{}\col{Q}(x) * x[f_1] * x[f_2]}{}$\\
$\colm{\theta} \leftarrow \colm{\theta_0}$\\
$\code{while(}\whilecond\code{)\{}$\\
$\hspace{0.2cm}\colm{\theta} = \dictbuild{f_1 \in \col{F}}{}(\colm{\theta}(f_1) - \summation{f_2 \in \col{F}}{}\colm{\theta}(f_2) * \col{M}(f_1)(f_2))$\\
$\code{\}}$\\
$\colm{\theta}$
\end{tabular}

\noindent Here \col{M} is the covar matrix (Section~\ref{sec:apps}) that 
is automatically introduced by static memoization and hoisted outside 
the loop thanks to loop-invariant code motion. However, for the cases where 
the number of features is more than the number of elements, none of this would 
happen. This is because the loop scheduling optimization is not applied,
and as a result static memoization and code motion will not be applied too.  
\end{example}

\subsection{Schema Specialization}
\label{sec:schemaspec}

At this stage, the dictionaries with statically-known keys of type \fieldtype{} 
(e.g., relations with the list of attributes provided as schema) are converted 
into records. Before checking for such dictionaries, 
partial evaluation transformations (Figure~\ref{fig:lang_opt_pe}) such as loop unrolling are performed.
Also, the dynamic field accesses are converted to static field accesses.
Figure~\ref{fig:lang_opt_sch} gives the schema specialization rules.

\begin{example}
	Schema specialization converts the dictionary encoding the $\colm{\theta}$ parameter into a record $\theta$ and all dynamic field accesses into static ones. The loop responsible for constructing the dictionary \col{M} is unrolled into the creation of the record $M$. The transformed \slang program is:

\noindent
\begin{tabular}{l}
$\letbinding{M}{\record{i=\record{...,c=\summation{x \in \dom{\col{Q}}}{}\col{Q}(x) \mult x.i \mult x.c,...},...}}{}$\\
$\theta \leftarrow \theta_0$\\
\end{tabular}

\begin{tabular}{l}
$\code{while(}\whilecond\code{)\{}$\\
$\hspace{0.2cm}\theta = \record{i=\theta.i - (...+\theta.c * M.i.c+\theta.p * M.i.p), ...}$\\
$\code{\}}$\\
$\theta$
\end{tabular}
\end{example}




\input{figures/lang-opts-form}

\subsection{Aggregate Query Optimizations}
\label{sec:lmfaoopt}

These optimizations focus on the query processing aspect of \lang programs and aim at interleaving the aggregates and joins to achieve faster computation and smaller result.

\begin{example}
In our running example, \col{Q} is the result of joining the input relations.
The computation of \col{Q} is represented as the following
\slang expression:

\noindent
\begin{tabular}{l}
\lett \col{Q} = \\
\summation{x_s \in \dom{\col{S}}}{} 
 \summation{x_r \in \dom{\col{R}}}{}
   \summation{x_i \in \dom{\col{I}}}{} $\big($\\
\tabt \letbind{k}{\record{$i=x_s.i,s=x_s.s,c=x_r.c,p=x_i.p$}}\\
\tabt \dict{k}{\col{S}($x_s$)\mult \col{R}($x_r$)\mult \col{I}($x_i$)\mult($x_s$.i==$x_i$.i)\mult($x_s$.s==$x_r$.s)}\\
$\big)$ 
\end{tabular}
\end{example}

We next explain how to compute the covar matrix $M$ over the query defining \col{Q} without materializing \col{Q}.

\smartpara{Extract aggregates} This pass analyzes the input \slang expression and extracts a batch of aggregates from it. 
These aggregates are then used in \slang expressions in lieu of 
those that rely on the materialization of the query result. In our example, $M$ captures such aggregates.

\smartpara{Join Tree Construction} This pass takes the join defining \col{Q} and produces a \textit{join tree}, where relations are nodes and an edge between two nodes is annotated with the variables on which the nodes of this edge join. The join order is computed using the state-of-the-art query optimization techniques~\cite{ioannidis1990randomized} and IFAQ assumes it is given as input. The join tree is used to factorize the computation of the aggregates in $M$.

\begin{example}
For the join of relations \col{S}, \col{R}, and \col{I} in our running example, we may consider the join tree \col{R}$\overset{i}{-}$\col{S}$\overset{s}{-}$\col{I} where \col{S} is the root and \col{R} and \col{I} are its children.
\end{example}

\smartpara{Aggregate Pushdown} This pass decomposes each aggregate into a 
view per edge in the join tree, resulting in a \textit{view tree}. 
Each
view is used to partially push down aggregates past joins and to allow the 
sharing of common views across the aggregate batch. 
The produced intertwining of aggregates and joins are injected back into the input \slang expression.

\begin{example}
Let us focus on the elements {\color{dgreen} $M_{c,p}$} and {\color{dred} $M_{c,c}$} of the (nested) record $M$ that defines aggregates:

\noindent
\begin{tabular}{l}
{\color{dgreen} \lett $M_{c,p}$ = 
\summation{x \in \dom{\col{Q}}}{}\col{Q}(x) \mult x.c \mult x.p
}\\
{\color{dred} \lett $M_{c,c}$ = 
\summation{x \in \dom{\col{Q}}}{}\col{Q}(x) \mult x.c \mult x.c
}\\
\lett M = \record{c=\record{...,
p={\textcolor{dgreen}{$M_{c,p}$}},
c={\textcolor{dred}{$M_{c,c}$}},...}
,...} \inn ...
\end{tabular}

\noindent Aggregate pushdown for {\textcolor{dgreen}{$M_{c,p}$}} yields the view tree: {\color{dgreen}$\colm{V_R}$---$M_{c,p}$---$\colm{V_I}$}.
The views {\color{dgreen}$\colm{V_R}$} and {\color{dgreen}$\colm{V_I}$} compute the aggregates for x.c and x.p while iterating over the relations \col{R} and \col{I}:

\noindent
\begin{tabular}{l}
{\color{dgreen} \lett $\colm{V_R}$ = 
\summation{x_r \in \dom{\col{R}}}{} \col{R}($x_r$) \mult \dict{\record{s=$x_r$.s}}{$x_r$.c} \inn } \\
{\color{dgreen} \lett $\colm{V_I}$ = 
\summation{x_i \in \dom{\col{I}}}{} \col{I}($x_i$) \mult \dict{\record{i=$x_i$.i}}{$x_i$.p} \inn }\\
\end{tabular}

\begin{tabular}{l}
{\color{dgreen} \lett $M_{c,p}$=%
\summation{x_s \in \dom{\col{S}}}{} \col{S}($x_s$) \mult $\colm{V_R}$(\record{s=$x_s$.s}) \mult $\colm{V_I}$(\record{i=$x_s$.i})}\\
\end{tabular}

\noindent A similar situation happens for the view tree 
{\color{dred}$\colm{V'_R}$---$M_{c,c}$---$\colm{V'_I}$}:

\noindent
\begin{tabular}{l}
{\color{dred} \lett $\colm{V'_R}$ = 
\summation{x_r \in \dom{\col{R}}}{} \col{R}($x_r$) \mult \dict{\record{s=$x_r$.s}}{$x_r$.c * $x_r$.c} \inn }\\
{\color{dred} \lett $\colm{V'_I}$ = 
\summation{x_i \in \dom{\col{I}}}{} \col{I}($x_i$) \mult \dict{\record{i=$x_i$.i}}{1} \inn }\\
{\color{dred} \lett $M_{c,c}$ = 
\summation{x_s \in \dom{\col{S}}}{} \col{S}($x_s$) \mult $\colm{V'_R}$(\record{s=$x_s$.s}) \mult $\colm{V'_I}$(\record{i=$x_s$.i})}\\
\end{tabular}
\end{example}

\noindent Computing these batches of aggregates requires multiple scans over each of the relations, the performance of which can be even worse than materializing the result of join.
Next, we show how to share the computation across these aggregates by fusing
the summations over the same collection.

\smartpara{Merge Views} This transformation consolidates the views generated in the previous transformation. All views computed at a node in the join tree will have the same key (the join variables shared between the node and its parent, or empty for the root node). These views are merged into a single view with the same key as the merged views and with all distinct aggregates of the merged views.

\smartpara{Multi-Aggregate Iteration} This transformation creates one summation over a relation for all aggregates of the view to be computed at that relation node. This computation seeks into the views computed at child nodes to fetch aggregate values used to compute the aggregates in the view. This transformation also shares computation and behaves similarly to horizontal loop fusion (Figure~\ref{fig:lang_opt_fuse}) on \slang expressions.

\begin{example}
	By performing horizontal loop fusion, one can merge {\color{dgreen}$\colm{V_R}$} and {\color{dred}$\colm{V'_R}$} into $\colm{W_R}$, as well as {\color{dgreen}$\colm{V_I}$} and {\color{dred}$\colm{V'_I}$} into $\colm{W_I}$:

\noindent
\begin{xtabular}{l}
\lett $\colm{W_R}$ = 
\summation{x_r \in \dom{\col{R}}}{} \col{R}($x_r$) \mult \\
\tabt\tabt\tabt\tabt\tabt \dict{\record{s=$x_r$.s}}{\record{{\color{dgreen}$v_R$=$x_r$.c}, {\color{dred}$v'_R$=$x_r$.c * $x_r$.c}}} \inn \\
\lett $\colm{W_I}$ = 
\summation{x_i \in \dom{\col{I}}}{} \col{I}($x_i$) \mult \\
\tabt\tabt\tabt\tabt\tabt \dict{\record{i=$x_i$.i}}{\record{{\color{dgreen}$v_I$=$x_i$.p}, {\color{dred}$v'_I$=1}}} \inn \\
\lett $M_{cc,pc}$ =
\summation{x_s \in \dom{\col{S}}}{} \col{S}($x_s$) \mult $\big($ \\
\tabt\tabt\tabt\tabt\tabt \lett $w_R$ = $\colm{W_R}$(\record{s=$x_s$.s}) \inn\\
\tabt\tabt\tabt\tabt\tabt \lett $w_I$ = $\colm{W_I}$(\record{i=$x_s$.i}) \inn\\
\tabt\tabt\tabt\tabt\tabt \record{${\color{dgreen}m_{c,p}=w_R.v_R\mult w_I.v_I},$
{\color{dred}$m_{c,c}=w_R.v'_R\mult w_I.v'_I$}} $\big)$ \inn\\
{\color{dgreen}\lett $M_{c,p}$ = $M_{cc,pc}.m_{c,p}$} \inn {\color{dred}\lett $M_{c,c}$ = $M_{cc,pc}.m_{c,c}$}
\end{xtabular}
\end{example}

Rather than iterating multiple times over each relation, here we thus
iterate over each relation only once. 

\smartpara{Dictionary To Trie}
This transformation pass converts relations and intermediate views from dictionaries into tries organized by join attributes.
Hence, the summations iterating over the full domain of a relation or view, are converted into 
nested summations over the domain of individual attributes. 
One key benefit of this transformation is more opportunities for factorizing the 
computation as well as hoisting the computation outside the introduced nested summations.

\begin{example}
As the intermediate views $\colm{W_R}$ and $\colm{W_I}$ have single fielded records as their
key, changing from a dictionary to a trie does not have any impact on them.
Thus, we focus only on the code for computing $M_{cc,pc}$. 
Rather than iterating over the domain of the key of the dictionary representing relation $\colm{S}$, 
we have to iterate in the key hierarchy of the converted trie structure, named as $\colm{S'}$.
This trie data-structure is a nested dictionary containing the domain of the 
field $s$ at its first level, and the domain of field $i$ at its second level.
The corresponding \slang expression is:

\noindent
\begin{tabular}{l}
\lett $M_{cc,pc}$ = \\
\tabt \summation{x_s \in \dom{\colm{S'}}}{} \\
\tabt\tabt \summation{x_i \in \dom{\colm{S'}(x_s)}}{} \\
\tabt\tabt\tabt $\colm{S'}$($x_s$)($x_i$) \mult $\big($ \\
\tabt\tabt\tabt\tabt\tabt \color{dblue}{\lett $w_R$ = $\colm{W_R}$(\record{s=$x_s$.s}) \inn}\\
\tabt\tabt\tabt\tabt\tabt \lett $w_I$ = $\colm{W_I}$(\record{i=$x_i$.i}) \inn\\
\tabt\tabt\tabt\tabt\tabt \record{$m_{c,p}=w_R.v_R\mult w_I.v_I,$\\
\tabt\tabt\tabt\tabt\tabt\tabt $m_{c,c}=w_R.v'_R\mult w_I.v'_I$} $\big)$ \inn
...
\end{tabular}

\noindent This expression can be further transformed using loop-invariant code motion 
(Figure~\ref{fig:lang_opt_licm}):

\noindent\begin{tabular}{l}
\lett $M_{cc,pc}$ = \\
\tabt \summation{x_s \in \dom{\colm{S'}}}{} \\
\tabt\tabt \color{dblue}{\lett $w_R$ = $\colm{W_R}$(\record{s=$x_s$.s}) \inn}\\
\tabt\tabt \summation{x_i \in \dom{\colm{S'}(x_s)}}{} \\
\tabt\tabt\tabt $\colm{S'}$($x_s$)($x_i$) \mult $\big($ \\
\tabt\tabt\tabt\tabt\tabt \lett $w_I$ = $\colm{W_I}$(\record{i=$x_i$.i}) \inn\\
\tabt\tabt\tabt\tabt\tabt \record{$m_{c,p}=w_R.v_R\mult w_I.v_I,$\\
\tabt\tabt\tabt\tabt\tabt\tabt $m_{c,c}=w_R.v'_R\mult w_I.v'_I$} $\big)$ \inn
...
\end{tabular}

\noindent Finally, the multiplication operands $w_R.v_R$ and
$w_R.v_R$ can be factored out of the inner summation:

\begin{tabular}{l}
\lett $M_{cc,pc}$ = \\
\tabt \summation{x_s \in \dom{\colm{S'}}}{} \\
\tabt\tabt \lett $w_R$ = $\colm{W_R}$(\record{s=$x_s$.s}) \inn\\
\tabt\tabt \color{dblue}{\record{$m_{c,p}=w_R.v_R,m_{c,c}=w_R.v'_R$}\mult} \\
\tabt\tabt \summation{x_i \in \dom{\colm{S'}(x_s)}}{} \\
\tabt\tabt\tabt $\colm{S'}$($x_s$)($x_i$) \mult $\big($ \\
\tabt\tabt\tabt\tabt\tabt \lett $w_I$ = $\colm{W_I}$(\record{i=$x_i$.i}) \inn\\
\tabt\tabt\tabt\tabt\tabt \record{$m_{c,p}=w_I.v_I,m_{c,c}=w_I.v'_I$} $\big)$ \inn 
...
\end{tabular}
\end{example}

Next, we move to more low-level optimizations including the physical representation of the
data structures.

\subsection{Data-Layout Synthesis}
\label{sec:dsopt}

\smartpara{Static Record Representation} The generated code for records can be dictionaries from 
the field name to the corresponding value. This is in essence what happens in the query interpreters of database
systems through using data dictionaries. However, as \slang uses code generation, this representation can be improved by generating static definitions for records (e.g., classes in Scala and structs in C/C++).

\smartpara{Immutable to Mutable} As \slang is a functional language, its 
data structures are immutable.
Even though such data structures improve the reasoning power, their runtime performance has room
for improvement. Especially, when summation produces collections, it can
lead to the production of many unused intermediate collections. 
This can be implemented more efficiently by appending the elements into a
mutable collection. 
As the summation construct is a tail recursive function, there are well-known optimization 
techniques~\cite{wadler1984listlessness} that can be applied to it;
an inital empty mutable collection at the beginning is allocated, and then at each iteration an 
in-place update is performed.

\smartpara{Scalar Replacement and Single-Field-Record Removal} In many cases, the intermediate records
are not \textit{escaped} from their scope, meaning that their fields can be treated 
as local variables. This optimization is called scalar replacement, and checking its applicability
is achieved using escape analysis. Furthermore, we have observed many cases where one is 
using a record with a single field. These records can be completely removed and substituted by their 
single field.

\smartpara{Memory Management} One major drawback of using languages with managed runtimes, such as JVM-based languages,
is the lack of control over the memory management. By using a low-level language such as C and C++ as the target
language, one can have more fine-grained control on memory management. This way, one can make sure that
the remaining records from the previous optimization, are mostly stack allocated rather than being
allocated on heap.

\smartpara{Dictionary to Array} \slang treats input relations as dictionaries from records to their multiplicity. However, in most cases the multiplicity is one, and 
it is more efficient to represent them as arrays. Furthermore, by statically
setting the multiplicities to one, the compiler sees more opportunities for constant folding and other optimizations.

\smartpara{Sorted Dictionary} In many cases, one iterates over a set of keys (coming from the domain of a dictionary) 
and looks for the associated values with that key in other dictionaries. 
If all these dictionaries are sorted on that key, the process of getting the associated value for a key 
becomes faster;
instead of looking for a key in the whole domain, it can ignore the already iterated domain.


%% file: figures/lang-opts-form.tex
\begin{figure*}[t]
\newcommand{\rewriteif}[1]{\color{gray4}{\textit{#1}}}
\begin{tabular}{cc}
\begin{minipage}{0.95\columnwidth}
\begin{tabular}{|p{0.37\columnwidth} p{0.095\columnwidth} p{0.38\columnwidth}|}
\hline
\exprind{1} \mult (\exprind{2} \add \exprind{3}) 
&\evalsto& 
\exprind{1} \mult \exprind{2} \add \exprind{1} \mult \exprind{3}
\\ \hline
\exprind{1} \mult \summation{x \in \exprind{2}}{\exprind{3}}
&\evalsto& 
\summation{x \in \exprind{2}}{(\exprind{1} \mult \exprind{3})}
\\ \hline
\exprind{1} \mult (-\exprind{2}) 
&\evalsto& 
-(\exprind{1} \mult \exprind{2}) 
\\ \hline
-\summation{x \in \exprind{2}}{\exprind{3}}
&\evalsto& 
\summation{x \in \exprind{2}}{-\exprind{3}}
\\ \hline
\end{tabular}
\vspace{-0.15cm}
\subcaption{Normalization Rules}
\label{fig:lang_opt_norm}
\begin{tabular}{|p{0.37\columnwidth} p{0.095\columnwidth} p{0.38\columnwidth}|}
\hline
\summation{x \in \exprind{1}}{\summation{y \in \exprind{2}}{\exprind{3}}}
&\evalsto& 
\summation{y \in \exprind{2}}{\summation{x \in \exprind{1}}{\exprind{3}}} \\ 
\multicolumn{3}{|c|}{\rewriteif{(\textit{if} $|\exprind{1}| > |\exprind{2}|$)}}
\\ \hline
\end{tabular}
\vspace{-0.15cm}
\subcaption{Loop Scheduling Rules}
\label{fig:lang_opt_loopsch}
\begin{tabular}{|p{0.37\columnwidth} p{0.095\columnwidth} p{0.38\columnwidth}|}
\hline
\exprind{1} \mult \exprind{2} \add \exprind{1} \mult \exprind{3}
&\evalsto& 
\exprind{1} \mult (\exprind{2} \add \exprind{3}) 
\\ \hline
\summation{x \in \exprind{2}}{(\exprind{1} \mult \exprind{3})} 
&\evalsto& 
\exprind{1} \mult \summation{x \in \exprind{2}}{\exprind{3}} \\
\multicolumn{3}{|c|}{\rewriteif{(\textit{if} x $\notin$ fvs(\exprind{1}))}}
\\ \hline
\end{tabular}
\vspace{-0.15cm}
\subcaption{Factorization Rules}
\label{fig:lang_opt_fact}
\begin{tabular}{|p{0.37\columnwidth} p{0.095\columnwidth} p{0.38\columnwidth}|}
\hline
\multirow{2}{*}{\summation{x \in \exprind{1}}{} \contexthole{\summation{y \in \exprind{2}}{\exprind{3}}}}
&\multirow{2}{*}{\evalsto}& 
\letbind{z}{\dictbuild{x \in \exprind{1}}{\summation{y \in \exprind{2}}{\exprind{3}}}}\\
& &
\summation{x \in \exprind{1}}{} \contexthole{z(x)}
\\ \hline
\end{tabular}
\vspace{-0.15cm}
\subcaption{Static Memoization Rules}
\label{fig:lang_opt_statmemo}
\begin{tabular}{|p{0.37\columnwidth} p{0.095\columnwidth} p{0.38\columnwidth}|}
\hline
\summation{x \in \exprind{1}}(\letbind{y}{\exprind{2}} \exprind{3})
&\evalsto& 
\letbind{y}{\exprind{2}}
\summation{x \in \exprind{1}}{\exprind{3}} \\
\multicolumn{3}{|c|}{\rewriteif{(\textit{if} x $\notin$ fvs(\exprind{2}))}}
\nextrule \\ \hline
x $\leftarrow$ \exprind{1} &\multirow{5}{*}{\evalsto}&  \letbind{y}{\exprind{3}}  \\
\while(\exprind{2}) & & x $\leftarrow$ \exprind{1} \\
\tabt \letbind{y}{\exprind{3}} && \while(\exprind{2}) \\
\tabt x $\leftarrow$ \exprind{4}& & \tabt x $\leftarrow$ \exprind{4}\\
\return x & & \return x
\\
\multicolumn{3}{|c|}{\rewriteif{(\textit{if} x $\notin$ fvs(\exprind{3}))}}
\\ \hline
\end{tabular}
\vspace{-0.15cm}
\subcaption{Loop-Invariant Code Motion Rules}
\label{fig:lang_opt_licm}
\end{minipage}
&
\begin{minipage}{1.05\columnwidth}
\centering
\begin{tabular}{|p{0.41\columnwidth} p{0.095\columnwidth} p{0.38\columnwidth}|}
\hline
\summation{x \in \set{\exprind{1},...,\exprind{n}}}{} \contexthole{x}
&\evalsto& 
\contexthole{\gind{e}{1}} + ... + \contexthole{\gind{e}{n}} \\ \hline
\dict{\exprind{1}}{\exprind{2}} + \dict{\exprind{1}}{\exprind{3}}
&\evalsto&
\dict{\exprind{1}}{\exprind{2} + \exprind{3}} \\ \hline
\dict{\exprind{1}}{\exprind{2}} + \dict{\exprind{3}}{\exprind{4}}
&\evalsto&
\dictmult{\exprind{1} \dictmapping \exprind{2}, \exprind{3} \dictmapping \exprind{4}} 
\\ \hline
\end{tabular}
\vspace{-0.15cm}
\subcaption{Partial Evaluation Rules}
\label{fig:lang_opt_pe}
\begin{tabular}{|p{0.41\columnwidth} p{0.095\columnwidth} p{0.38\columnwidth}|}
\hline
\fieldreflect{\exprind{1}}{\fieldlit{f}} & \evalsto& \exprind{1}.f
\nextrule\\ \hline
\dictmult{...,\fieldlit{\gind{f}{i}} \dictmapping \exprind{i},...}
&\evalsto&
\record{...,\gind{f}{i} = \exprind{i},...}
\nextrule\\ \hline
\exprind{1}(\exprind{2})
&\evalsto&
\fieldreflect{\exprind{1}}{\exprind{2}} \\ 
\multicolumn{3}{|c|}{\rewriteif{(\textit{if \exprind{1} is transformed})}}
\nextrule\\ \hline
\dictbuild{x \in \text{\set{...,\fieldlit{\gind{f}{i}},...}}}{} \contexthole{\fieldreflect{\exprind{1}}{x}}
&\evalsto& 
\record{...,\gind{f}{i} = \contexthole{\exprind{1}.\gind{f}{i}},...}
\\ \hline
\end{tabular}
\vspace{-0.15cm}
\subcaption{Schema Specialization Rules}
\label{fig:lang_opt_sch}
\begin{tabular}{|p{0.41\columnwidth} p{0.095\columnwidth} p{0.38\columnwidth}|}
\hline
\letbind{x}{\summation{z \in \exprind{1}}{\exprind{2}}} 
&\multirow{3}{*}{\evalsto}& 
\lett{} t =
\\
\letbind{y}{\summation{z \in \exprind{1}}{\exprind{3}}} & &
\tabt \summation{z \in \exprind{1}}{\record{x=\exprind{2},y=\exprind{3}}}
 \\[-0.125em]
\contexthole{x, y} & & \inn{} \contexthole{t.x, t.y}
\\ \hline
\end{tabular}
\vspace{-0.15cm}
\subcaption{Loop Fusion Rule}
\label{fig:lang_opt_fuse}
\begin{tabular}{|p{0.41\columnwidth} p{0.095\columnwidth} p{0.38\columnwidth}|}
\hline
\lett{} x = \exprind{0} \inn{} \contexthole{x}
&\evalsto& 
\contexthole{\exprind{0}}
\nextrule\\ \hline
\lett{} x = \exprind{0} \inn{} \exprind{1}
&\evalsto& 
\exprind{1} \\
\multicolumn{3}{|c|}{\rewriteif{(if x $\not\in$ fvs(\exprind{1}))}}
\nextrule\\ \hline
\lett{} x = & & \lett{} y = \exprind{0} \inn{}\\
   \tabt
  \lett{} y = \exprind{0} \inn{} \exprind{1} &\evalsto& \lett{} x = \exprind{1} \\
\inn{} \exprind{2}
& & 
\inn{} \exprind{2} 
\nextrule\\ \hline
\lett{} x = \exprind{0} \inn{} & & \lett{} x = \exprind{0} \inn{} \\
\lett{} y = \exprind{0} \inn{} &\evalsto& \contexthole{x, x} \\
\contexthole{x, y} & & 
\nextrule\\ \hline
\end{tabular}
\vspace{-0.15cm}
\subcaption{Generic Optimization Rules}
\label{fig:lang_opt_gen}
\end{minipage}
\\ 
\end{tabular}
\vspace{.10cm}
\Description{Transformation Rules of \lang{}. We use \contexthole{\exprind{1}} on the left-hand-side to denote a context in which \exprind{1} is used, and \contexthole{\exprind{2}} on the right-hand-side represents the same context where each occurrence of \exprind{1} is substituted by \exprind{2}.}
\caption{Transformation Rules of \lang{}. We use \contexthole{\exprind{1}} on the left-hand-side to denote a context in which \exprind{1} is used, and \contexthole{\exprind{2}} on the right-hand-side represents the same context where each occurrence of \exprind{1} is substituted by \exprind{2}.}
\label{fig:lang_opts}
\end{figure*}


%% file: exp.tex
\begin{table}[t]
  \centering
  \caption{Characteristics of the Retailer and Favorita datasets.}

  \begin{tabular}{|l|r|r|}\hline
    & \multicolumn{1}{|c|}{ Retailer }
    & \multicolumn{1}{|c|}{ Favorita }\\\hline
    Tuples/Size of Database     & 87M(1.5GB)   & 125M(2.5GB)  \\
    Tuples/Size of Join Result  & 86M(17GB)  & 127M(2.6GB)  \\
    Relations/Continuous Attrs & 5 / 35   & 5 / 6    \\\hline
  \end{tabular}
  \label{table:datasetstats}
\end{table}

\begin{figure*}[t]
  \begin{subfigure}[b]{0.49\textwidth}
    \includegraphics[width=\textwidth]{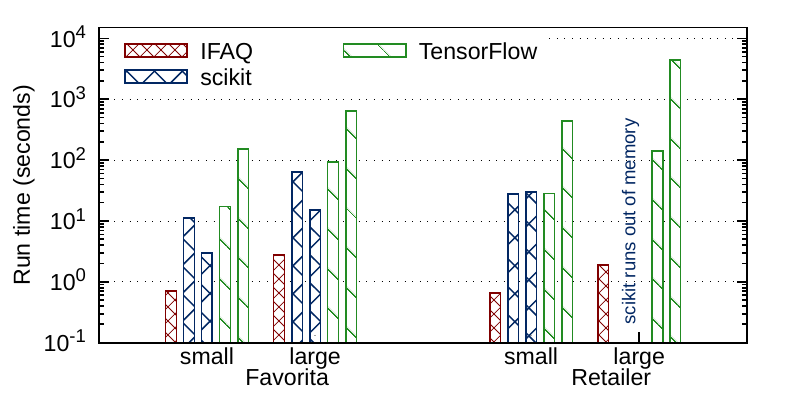}
  \end{subfigure}\hfill
  \begin{subfigure}[b]{0.49\textwidth}
    \includegraphics[width=\textwidth]{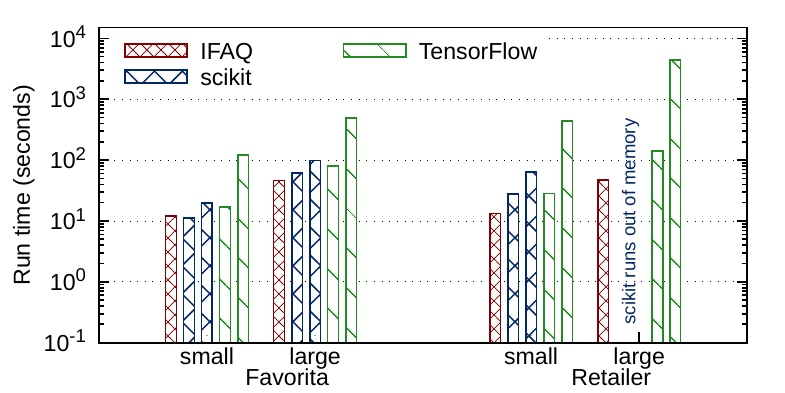}
  \end{subfigure}
  \Description{Performance comparison of \lang, TensorFlow, and scikit-learn for
    learning linear regression models and regression trees.}
  \caption{Performance comparison of \lang, TensorFlow, and scikit-learn for
    learning linear regression models (left) and regression trees (right) over
    the two variants of Favorita and Retailer. For TensorFlow and scikit-learn,
    the performance is separated into the time it takes to compute the training
    dataset (left bar) and the time for learning the model (right bar). }
 \label{fig:evale2e}
\end{figure*}

\section{Experimental Results}
\label{sec:exp}
In this section, we benchmark \lang, TensorFlow and scikit-learn for learning
linear regression and regression tree models using two real-world datasets. We
then micro benchmark individual \lang optimizations.

\smartpara{Datasets} We consider two real-world datasets, used in retail
forecasting scenarios (Table~\ref{table:datasetstats}): (1)
\emph{Favorita}~\cite{favorita} is a publicly available Kaggle dataset; and (2)
\emph{Retailer} is is a dataset from a US
retailer~\cite{Schleich:2016:LLR:2882903.2882939}. Both datasets have a fact
table with information about sales and respectively inventory for different
stores, dates, and products. The other tables provide information about the
products and stores.  We consider common retail-forecasting models that predict
future sales for Favorita, and future inventory demand for Retailer. For each
dataset, we learn the models over the natural join of all relations, which
include the sales and respectively inventory for all dates except the last
month.  The sales and respectively inventory for the last month is used as the
test dataset to measure model accuracy. We also consider a smaller variant for
each dataset whose size is 25\% of the join result.

\smartpara{Experimental Setup} All experiments were performed on an
Intel(R) Core(TM) i7-4770/3.40GHz/32GB/Ubuntu
18.04. We use g++ 6.4.0 for compiling the generated C++ code using the O3
optimization flag.  For all the experiments, we compute the average of four
subsequent runs.  We do not consider the time to load the database into RAM and
assume that all relations are indexed by their join attributes.

We learn the models over all continuous attributes for Favorita and Retailer. 
We learn regression trees up to depth four (i.e., max 31 nodes).


The input to \lang is a program that performs batch gradient
descent optimization for linear regression models or the CART algorithm for
learning regression trees. \lang automatically
optimizes the code. For linear regression, it employs the full suite
of optimizations, including the memoization and hoisting of the covar
matrix. For regression trees, the aggregates cannot be hoisted, but they still
benefit from the lower level optimizations, including loop fusion and
data-layout synthesis. The optimized code interleaves the computation
of the training dataset with the learning task.

TensorFlow learns the linear regression model with the predefined
LinearRegressor estimator. We report the time to perform one epoch with a batch
size of 100,000 instances. This batch size gave the best performance/accuracy
trade-off. The regression trees are learned with the BoostedTrees
estimator. Since TensorFlow and scikit-learn do not support the query processing
task, we use Pandas DataFrames to materialize the training
dataset~\cite{mckinney2012python}. We compare the accuracy of all systems by
comparing the root-mean-square-error (RMSE) over the test dataset. For linear
regression, we compare against the closed-form solution of scikit-learn or
MADlib~\cite{hellerstein2012madlib} when scikit-learn fails.

Scikit-learn requires that the training dataset is represented in memory, which
can lead to out-of-memory errors.
In our experiments, TensorFlow also runs out of memory for
learning over the full Retailer dataset. In this case, we wrote the data to a
file and used the CSV parser to iteratively load mini-batches of data during
learning. We do not include the time for writing the dataset to the file.

We attempted to benchmark against mlpack~\cite{mlpack2018}, an efficient C++
machine learning library, but it runs out-of-memory on all our experiments. One
explanation is that mlpack copies the input data to compute its transpose. In
our experiments, mlpack failed for as little as 5\% of the Favorita training
dataset (100MB on disk).

\begin{figure*}[t!]
\centering
\begin{minipage}{0.49\textwidth}
\includegraphics[width=\columnwidth]{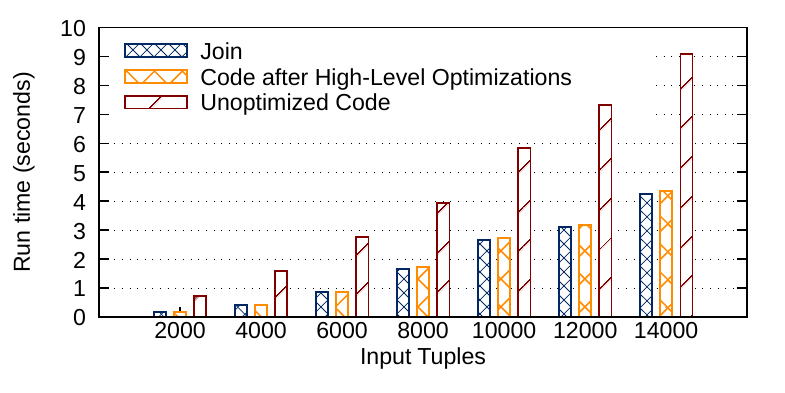}
\end{minipage}
\begin{minipage}{0.49\textwidth}
\includegraphics[width=\columnwidth]{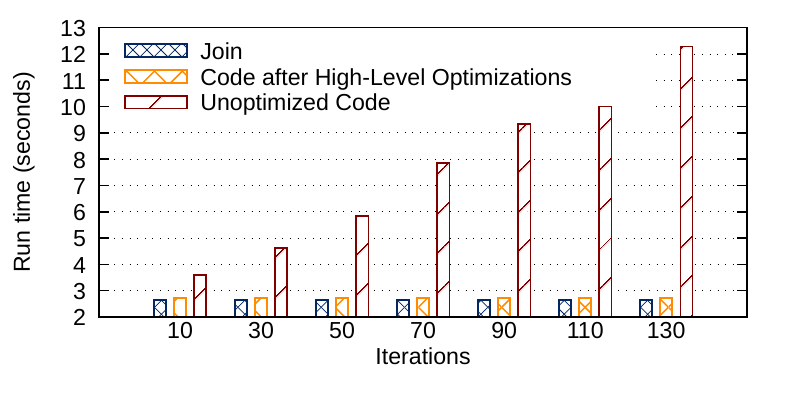}
\end{minipage}
\Description{Impact of high-level optimizations for learning the linear regression model
using BGD}
\caption{Impact of high-level optimizations for learning the linear regression model
using BGD by varying the number of input tuples for 50 iterations (left) and 
varying the number iterations for 10,000 input tuples (right). The join computation is identical for both unoptimized and optimized code and shown by a separate bar.}
\label{fig:evalindhl}
\end{figure*}

\begin{figure*}[t!]
\centering
\begin{minipage}{0.49\textwidth}
\includegraphics[width=\columnwidth]{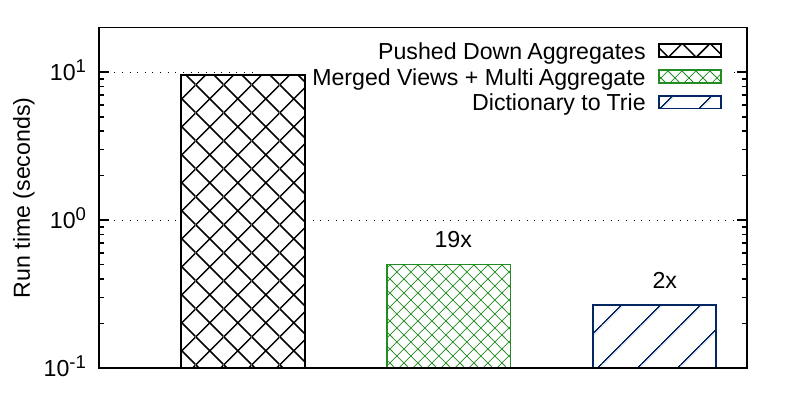}
\subcaption{Aggregation optimizations.}
\label{fig:evalindcovar1}
\end{minipage}
\begin{minipage}{0.49\textwidth}
\includegraphics[width=\columnwidth]{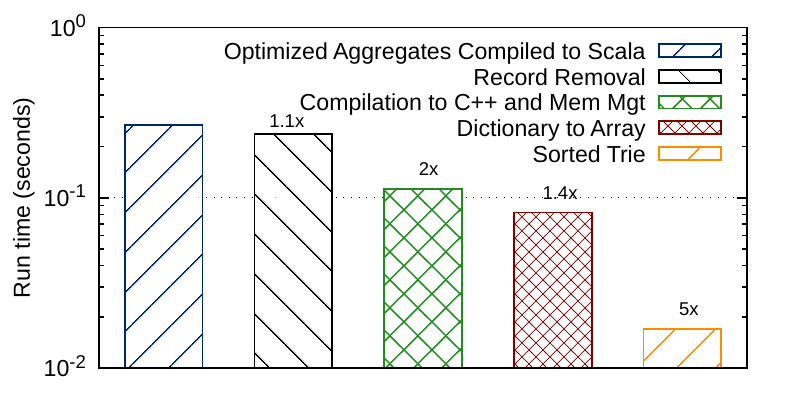}
\subcaption{Low-level optimizations.}
\label{fig:evalindcovar2}
\end{minipage}
\Description{Impact of aggregation and low-level optimizations for the covar matrix
  computation.}
\caption{Impact of aggregation and low-level optimizations for the covar matrix
  computation for 1,000,000 input tuples.}
\label{fig:evalindcovar}
\end{figure*}

\smartpara{Benchmarking End-To-End Learning} Figure~\ref{fig:evale2e} shows the
results of the end-to-end benchmarks for learning linear regression and
regression tree models in \lang, scikit-learn, and TensorFlow. The runtime for
the latter two is the sum of two components, depicted in Figure 6
by two bars: (1) the left bar is the time to materialize the
project-join query that defines the training dataset; 
(2) the right bar is the time to learn the model. \lang learns the model
directly over the input database in one computation step.

\lang significantly outperforms both scikit-learn and TensorFlow, whenever they do not fail. The end-to-end computation in \lang is consistently faster than it takes the competitors to materialize the training dataset. It is remarkable that IFAQ outperforms them even though it runs in one thread whereas the competitors use 8 threads (one per CPU core) for learning the model.

The reason for the performance improvement is due to the suite of optimizations in \lang. First, \lang represents the core computation of the learning algorithms as aggregate
queries and computes them directly over the input database, which is
significantly smaller than the join result (Table~\ref{table:datasetstats}). It
then uses several optimizations to further improve the computation of these
aggregates (Section~\ref{sec:opts}). We evaluate the impact of individual
optimizations below.

For linear regression, the RMSE for \lang is within 1\% of the closed form
solution. TensorFlow computes a single epoch, and the RMSE is worse than \lang's
(e.g., for Retailer small it is 3\% higher).  TensforFlow would thus need more
epochs to achieve the accuracy of \lang. Scikit-learn and \lang learn very similar
regression trees (using CART variants) so the accuracies are very close.

\smartpara{Categorical Attributes}
The previous experiments only report on runtime performance for datasets with continuous 
features only. All aforementioned systems, including \lang, can also support categorical 
features via one-hot encoding. This translates into 87 features for Retailer and 526 
features for Favorita, one per category of each categorical attribute and one per 
continuous attribute. We found that this encoding leads to an out-of-memory error for 
scikit-learn. Furthermore, \lang becomes much slower as it needs to 
generate quadratically many aggregates in the number of features. We leave 
the more efficient handling of categorical attributes, e.g., by extending \lang 
with sparse tensors as in LMFAO~\cite{Schleich:2019:LAE:3299869.3324961}, as future work.

\smartpara{Compilation Overhead} The gcc compilation of the generated C++
code for Retailer is 
on average in 4.3 and 8.3 seconds for linear regression and respectively
regression trees, while for Favorita this is 9.7 and respectively 2.4 seconds.

\smartpara{Individual Optimizations} We evaluate the effect of individual
optimizations for learning the linear regression model using BGD on a subset of
the Favorita dataset. We start with the impact of high-level optimizations using
an interpreter for \dlang.  Then, we continue with the impact of the aggregate
query optimizations on an interpreter for \slang. Finally, we show the impact
of lower level optimizations on a compiler for \slang which generates Scala or C++
code.

To evaluate the effect of individual high-level optimizations, we start with an
unoptimized program that first materializes the join and then learns the model on
this materialized dataset. 
This is what existing ML tools do. 
For linear regression, applying high-level optimizations
results in hoisting the covar matrix out of the while loop
(Section~\ref{sec:hlopt}).

Figure~\ref{fig:evalindhl} shows the impact of high-level optimizations by
showing the performance of unoptimized and optimized programs. Both these
programs need to materialize the join result. However, the unoptimized program
needs to perform aggregations at each iteration on the training dataset, whereas
in the optimized program, most of these data-intensive computations are hoisted
outside the loop.  The two graphs show that most of the computation for the
optimized program is dedicated to the join computation. In addition, the
increase in the number of iterations has a negligible impact on the performance
of the optimized program.

For the following optimization layers, we focus on the data-intensive
computation, i.e., computing the covar matrix.

The impact of aggregate optimizations (Section~\ref{sec:lmfaoopt}) on the
performance of computing the covar matrix is shown in
Figure~\ref{fig:evalindcovar1}:
(i) merging views and multi-aggregate iteration have a significant impact
on the performance thanks to horizontal loop fusion; (ii) converting dictionaries 
to tries also improves the performance
thanks to the sharing of computation enabled by factorization.

Finally, Figure~\ref{fig:evalindcovar2} shows the impact of low-level
optimizations (Section~\ref{sec:dsopt}) on the computation of the covar
matrix. As the code is more optimized, we can process a significantly larger
amount of data. The leftmost bar shows the code after aggregate optimizations
and compilation in Scala. The following two optimizations have the highest
impact: (i) generating C++ code with efficient memory management has 2x
performance improvement, which requires less heap allocation that the Scala
code; and (ii) using an already sorted trie rather than a hash-table trie
data-structure gives 5x speedup.


%% file: related.tex
\section{Related Work}
\label{sec:rel}




\smartpara{High Performance Computing}\ There is high demand for efficient matrix processing in numerical and scientific computing. BLAS~\cite{Dongarra1990} exposes a set of low-level routines for common linear
algebra primitives used in higher-level libraries including LINPACK,
LAPACK, and ScaLAPACK for parallel processing.
Highly optimized BLAS implementations are provided for dense linear algebra by hardware vendors such as Intel or AMD and code generators such
as ATLAS~\cite{Whaley1999} and for sparse linear algebra by Combinatorial BLAS~\cite{bulucc2011combinatorial}. 
HPAT~\cite{totoni2017hpat} compiles a high-level
scripting language into high-performance parallel code. \lang can reuse
techniques developed in all the aforementioned tools.

\smartpara{Numerical and Linear Algebra DSLs} The compilers of high-level linear
algebra DSLs, such as Lift~\cite{Steuwer:2015:GPP:2784731.2784754},
Opt~\cite{devito2017opt}, Halide~\cite{ragan2013halide},
Diderot~\cite{chiw2012diderot}, and OptiML~\cite{sujeeth2011optiml}, focus on
generating efficient parallel code from the high-level programs.

Spiral~\cite{spiral}
provides a domain-specific compiler for synthesizing Digital Signal Processing kernels, e.g., Fourier transforms.  It is based on the SPL~\cite{Xiong:2001:SLC:378795.378860} language that expresses recursion and mathematical formulas. 
The LGen~\cite{Spampinato:2014:BLA:2581122.2544155} and  SLinGen~\cite{Spampinato:2018:PGS:3179541.3168812} systems target small-scale  computations involving fixed-size linear algebra expressions common in graphics and media processing applications. They use two-level DSLs, namely LL to perform tiling decisions and $\Sigma$-LL to enable loop level optimizations. The generated output is a C function that includes intrinsics to enable SIMD vector extensions.

TACO~\cite{Kjolstad:2017:TAC:3152284.3133901} generates efficient low-level code 
for compound linear algebra operations on dense and sparse matrices.
Workspaces~\cite{Kjolstad:2019:TAC:3314872.3314894} further improve the performance of the generated kernels by increasing sharing, and enabling optimizations such as loop-invariant code motion. It has similarities to our static memoization optimization (Section~\ref{sec:hlopt}).

APL~\cite{iverson1962programming} is the pioneer of array languages, 
the design of which inspired functional array languages such
as SAC~\cite{Grelck2006}, Futhark~\cite{henriksen2017futhark}, and $\widetilde{\text{F}}$~\cite{shaikhha2019efficient}. The key feature of all these languages
is the support for fusion~\cite{Svensson:2014:DPA:2636228.2636231,edsl-push,Claessen:2012:EAC:2103736.2103740,dps_fhpc}, which is essential for efficient low-level code generation.

Finally, there are mainstream programming languages used by data scientists.
The R programming language~\cite{team2013r} is widely used by statisticians and data miners. It provides a standard language for statistical computing that includes arithmetic, array manipulation, object oriented programming and system calls. Similarly, MATLAB provides a wide range of toolboxes for data analytics tasks.

These DSLs do not express the hybrid query processing and
machine learning workloads that \lang supports. Next, we highlight techniques used to
overcome this issue.

\smartpara{Functional Languages for Data Processing}
Apart from the linear-algebra-based DSLs, there are languages for data processing based on the nested relational model~\cite{roth1988extended} and the monad calculus and monoid comprehension~\cite{monad-calc-1, monad-calc-2, monad-comprehension, query-comprehension, query-comprehension-2, monoid-comprehension, wong2000kleisli, Buneman:1995:PPC:210500.210501}. 
Functional collection programming abstractions existing in the standard library of the mainstream programming languages 
such as Scala, Haskell, and recently Java 8 are also based on these languages. 
A similar abstraction is used in distributed data processing frameworks such as Spark~\cite{rdd},
as well as as an intermediate representation in Weld~\cite{palkar2017weld}.
Similar to \lang, DBToaster~\cite{DBToaster:VLDBJ2014} uses a bag-based collection, yet for incremental view maintenance.
A key challenge for such collection-based languages is efficient code generation, which has been investigated both by the DB~\cite{palkar2017weld,dblablb,legobase_tods} and PL~\cite{Kiselyov:2017:SFC:3009837.3009880,Mainland:2013:EVI:2500365.2500601,jfppushpull} communities.
The \lang languages are also inspired by the same model, with more similarities to monoid comprehension; the summation operator can be considered as a monoid comprehension.
However, they are more expressive and allow for a wider range of optimizations and generation of efficient low-level code for aggregate batches.

\smartpara{In-Database Machine Learning}
The common approach to learning models over databases is to first construct the training dataset using a query engine, and then learn the model over the materialized training dataset using a machine learning framework. Python Pandas or database systems such as PostgreSQL and SparkSQL~\cite{Armbrust:2015:SSR:2723372.2742797} are the most common tools used for the first step. The second step commonly uses scikit-learn~\cite{pedregosa2011scikit}, TensorFlow~\cite{abadi2016tensorflow}, PyTorch~\cite{paszke2017automatic}, R~\cite{team2013r}, MLlib~\cite{Meng:2016:MML:2946645.2946679}, SystemML~\cite{ghoting2011systemml}, or XGBoost~\cite{chen2016xgboost}. 
On-going efforts avoid the expensive data export/import at the interface between the two steps, e.g., MLlib over SparkSQL, and the Python packages over Pandas. 
MADlib~\cite{hellerstein2012madlib}, Bismarck~\cite{Feng:2012:TUA:2213836.2213874}, and GLADE PF-OLA~\cite{qin2015speculative}  define ML tasks as user-defined aggregate
functions inside database systems so that the ML tasks share the same processing environment with the query engine. The two steps are nevertheless treated as black boxes and executed after the training dataset is materialized.

Alternative approaches avoid the materialization of the training dataset. Morpheus~\cite{chen2017towards,li2019enabling} reformulates in-database machine learning tasks in terms of linear algebra operations on top of R~\cite{chen2017towards} and NumPy~\cite{li2019enabling}. It supports a limited class of schema definitions (i.e., key-foreign key star or chain joins) and learns models expressible in linear algebra. IFAQ is the latest development of a sustained effort to train efficiently machine learning models over arbitrary joins. Its predecessors are: F~\cite{Olteanu:2016:FD:3003665.3003667, Schleich:2016:LLR:2882903.2882939} for  linear regression; AC/DC~\cite{Khamis:2018:AIL:3209889.3209896} for polynomial regression and factorization machines; and LMFAO~\cite{Schleich:2019:LAE:3299869.3324961} for models whose data-intensive computation can be expressed as batches of aggregates. IFAQ introduces: high-level optimizations, a systematic treatment of the various optimizations through compilation stages, and languages expressive enough to capture the full ML application.


%% file: concl.tex
\section{Conclusion and Future Work}
\label{sec:concl}

In this paper we introduced \lang, an optimization framework for relational learning applications, which includes the construction of the data matrix via feature extraction queries over multi-relational data and the subsequent model training step. \lang takes as input such an application expressed in a dynamically-typed language that captures a fragment of scripting languages such as Python. \lang then applies multiple layers of optimizations that are inspired by techniques developed by the databases, programming languages, and high-performance computing communities. As proof of concept, we showed that IFAQ outperforms mainstream approaches such as TensorFlow and Scikit by orders of magnitude for linear regression and regression tree models.

This work opens exciting avenues of future research. New technical developments include: a compilation approach to capture LMFAO's efficient support for categorical variables with multi-root join trees and group-by aggregates~\cite{Schleich:2019:LAE:3299869.3324961}; support for parallelization and many-core architectures; and an investigation of the trade-off between runtime performance and size of generated C++ code for models with high degree and many parameters (e.g., factorization machines). We would also like to improve the usability of IFAQ as follows: build an IFAQ library of optimization algorithms and ML models beyond the simple ones discussed in this paper and including boosting trees, random forests, and neural networks; generate optimized code for model selection over different subsets of the given variables; allow IFAQ to work directly on Jupyter notebooks that specify the construction of the data matrix and the model training; and investigate whether the IFAQ compilation techniques can be incorporated into popular data science tools such as Scikit and TensorFlow.